\documentclass[reprint, superscriptaddress]{revtex4-1}

\usepackage{amsmath}
\usepackage{graphicx}

\begin{document}

\title{Retrieving broadband ultrashort THz wave packets through lossy channels}

\author{A.~D.~Koulouklidis}
\affiliation{Institute of Electronic Structure and Laser (IESL), Foundation for
             Research and Technology - Hellas (FORTH), P.O. Box 1527, GR-71110
             Heraklion, Greece}
\affiliation{Materials Science and Technology Department, University of Crete,
             71003, Heraklion, Greece}

\author{V.~Yu.~Fedorov}
\affiliation{Institute of Electronic Structure and Laser (IESL), Foundation for
             Research and Technology - Hellas (FORTH), P.O. Box 1527, GR-71110
             Heraklion, Greece}
\affiliation{P.~N.~Lebedev Physical Institute of the Russian Academy of
             Sciences, 53 Leninskiy Prospekt, 119991, Moscow, Russia}
\affiliation{Science Program, Texas A\&M University at Qatar, P.O. Box 23874
             Doha, Qatar}

\author{S.~Tzortzakis}
\affiliation{Institute of Electronic Structure and Laser (IESL), Foundation for
             Research and Technology - Hellas (FORTH), P.O. Box 1527, GR-71110
             Heraklion, Greece}
\affiliation{Materials Science and Technology Department, University of Crete,
             71003, Heraklion, Greece}
\affiliation{Science Program, Texas A\&M University at Qatar, P.O. Box 23874
             Doha, Qatar}
\email{stzortz@iesl.forth.gr}

\date{\today}

\begin{abstract}
Most transmission and detection channels fail to faithfully support broadband
wave packets because of physical limitations, like chromatic dispersion and
absorption.
We explore the case of lossy detection of ultrashort THz pulses using the
widespread electro-optic detection scheme.
We demonstrate that one can fully recover the original THz pulse shape, duration
and amplitude, using a simple experimental procedure and a reconstruction
algorithm which encodes the physical properties of the detection system.
\end{abstract}

\maketitle

\section*{Introduction}
Nowadays electro-magnetic fields in the THz frequency range are of interest for
a wide range of applications, including spectroscopy, nondestructive testing,
tomographic imaging, and many others~\cite{Tonouchi2007,Ferguson2002,
Jewariya2013,Massaouti2013,Zhao2014}.
Modern THz sources are able to generate THz radiation with intensities high
enough to observe nonlinear effects~\cite{Turchinovich2012,Zalkovskij2013,
Hebling2009}.
To date there are two major techniques for the generation of intense THz pulses
on table-top setups: optical rectification in nonlinear
crystals~\cite{Huang2014} and two-color filamentation (photoionization of gases
by symmetry-broken laser fields)~\cite{Kim2008,Oh2013}.
Optical rectification allows one to generate THz pulses with very high energies
(up to 900\,$\mu$J) and spectral widths up to 5\,THz~\cite{Vicario2014}.
In turn, THz pulses generated by two-color filamentation have less energy (up to
7\,$\mu$J) but their spectrum is much broader, ranging from almost zero up to
approximately 60\,THz~\cite{Oh2013b}.

Although the generation of intense broadband THz radiation is a challenging
task, its detection can be even more complicated.
In the majority of THz applications we need a way to coherently detect a pulsed
THz wave with a simultaneous measurement of its amplitude and phase.
The most commonly used coherent detectors for THz radiation are photoconductive
antennas~\cite{Brener1996,Cai1997} and electro-optic crystals~\cite{Wu1995,
Nahata1996,Kubler2005}.
However, they have a limited detection bandwidth due to a finite carrier
lifetime in antenna photocunductors and phonon resonances in electro-optic
crystals.
A typical THz pulse emitted during two-color filamentation has a spectrum which
extends far beyond the detection limits of these two types of detectors.
To date, the most broadband method for coherent detection of THz radiation is
Air Biased Coherent Detection (ABCD)~\cite{Dai2011,Lu2011,Lu2012}.
However the ABCD method is less sensitive than electro-optic detectors
especially at low repetition rates, while its physical limitations are not fully
studied.
As a consequence, the majority of laboratories still use the electro-optic
crystals as the main tool for coherent detection of THz radiation.

Since electro-optic detectors can measure only a small part of the generated THz
spectrum, the detected THz electric field is strongly distorted.
As a result, during electro-optic detection lots of information about the
spectral content, shape and duration of the original THz pulse is lost.
Although it may seem that this information is lost irrevocably, in this work we
show how using a small number of realistic assumptions together with a simple
experimental procedure one can retrieve the full information about the original
THz wave packet and restore its parameters.
Our reconstruction method is based on the nonlinear curve fitting of a recorded
electro-optic signal.
However, in contrast to the standard nonlinear curve fitting where the fitting
function is expressed in a closed-form, our fitting function implements a
numerical simulation of electro-optic detection.
This method is accompanied by an experimental procedure where we stretch the
input laser pulse in order to increase the duration of the generated THz pulses
up to the point where their spectrum becomes narrower than the detection
bandwidth of the electro-optic crystal.
At this point we restore parameters of the generated THz pulses and
appropriately extrapolate them to the region of the original THz pulse duration.
As a result, we are able to retrieve the shape, duration, and absolute amplitude
of the original THz pulse using only the partial information provided by the
electro-optic detection.

\section{The effect of a limited bandwidth}
At first let us demonstrate how the limited bandwidth of a detector affects our
perception of a THz pulse.

In our experiments we generate THz pulses using the two-color filamentation
scheme.
Our Ti:sapphire chirped-pulse amplification laser system generates 39\,fs (FWHM)
pulses, at 800\,nm central wavelength and maximum energy 2.3\,mJ.
We focus the input laser pulse by a lens with 200\,mm focal length,
followed by a 50\,$\mu$m thick $\beta$-barium borate (BBO) Type-I crystal cut at
$29.9^\circ$ angle, which generates the second harmonic.
After the BBO crystal, at the focus of the lens, the two-color pulses create a
plasma filament that emits THz radiation.
In order to avoid the absorption in water vapor we place our setup into a purge
gas chamber filled with dry air.
Using a pair of parabolic mirrors we gather the emitted THz radiation in the far
field and send it to a THz detector.
To detect the generated THz pulses we apply either electro-optic detection or
ABCD measurements.
For electro-optic detection we use 0.5\,mm zinc telluride (ZnTe) and 100\,$\mu$m
gallium phosphide (GaP) crystals.

In addition to electro-optic and ABCD measurements we also conducted
measurements of the generated THz pulses using Michelson interferometry as a
reference method.
We should stress that although originally Michelson interferometry is not a
method of coherent detection (it cannot measure explicitly the phase of a THz
pulse), it allows us though to detect the whole THz spectrum without
limitations.

\begin{figure}
    \includegraphics{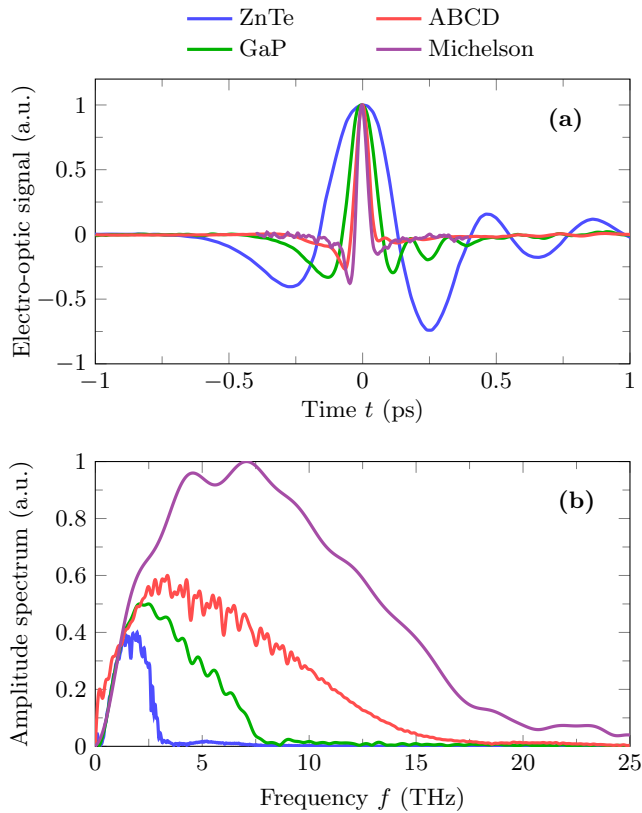}%
    \caption{\label{fig:compare_detectors}%
             Normalized THz electro-optic signals (a) and their amplitude
             spectra (b) for a THz pulse measured by different techniques,
             namely, by ZnTe and GaP electro-optic crystals, as well as by the
             ABCD method and Michelson interferometry (the spectra are
             normalized appropriately for clarity).}
\end{figure}

Figure~\ref{fig:compare_detectors} shows the signals and their amplitude spectra
obtained by the different detection techniques.
We see that the shortest spectrum is detected by the ZnTe crystal.
It is followed by spectra measured by the GaP crystal and then by the ABCD
measurement.
The widest spectrum is registered by Michelson interferometry.
The detection bandwidth of electro-optic crystals is mainly restricted by their
transverse optical phonon resonances.
In ZnTe crystals the first phonon resonance is located at 5.3\,THz and in GaP at
11\,THz~\cite{Casalbuoni2008}.
In turn, the detection bandwidth of the ABCD method can be estimated as the
inverse duration of the probe pulse (the main limiting parameter).
In our experiments, the duration of the probe pulse is about 70\,fs (the probe
pulse is longer than the input pulse due to the dispersion in the guiding
optics) which corresponds to a detection bandwidth of about 14\,THz.

In Fig.~\ref{fig:compare_detectors} we also see that the signals obtained by the
different detection techniques have different durations and shapes.
To be more quantitative, let us define the duration $\tau_s$ of the detected
signal as the duration of its first oscillation, i.e., the time between the
first minimum and the first maximum of the signal.
Using this definition, we find that the durations $\tau_s$ of the signals in
Fig.~\ref{fig:compare_detectors} are the following: 275\,fs for the ZnTe
crystal, 127\,fs for the GaP crystal, and 66\,fs for the ABCD method.
In principle, the signal from the Michelson interferometer does not reproduce
the electric field of the incoming THz pulse.
However, the period of the Michelson interferogram corresponds to such an
optical path difference that the temporal delay between the pulses in the two
arms of the Michelson interferometer is equal to the period of the detected THz
electric field.
Therefore, taking into account our definition of $\tau_s$, we can use the
Michelson interferogram in order to estimate the duration of the generated THz
pulses.
In Fig.~\ref{fig:compare_detectors} we see that the signal obtained by the
Michelson interferometer shows a duration $\tau_s$ of 44\,fs, which we assume
to be the closest estimate of the actual THz pulse duration.

Thus, even using the best of our electro-optic crystals we measure the THz pulse
duration being almost three times larger than its actual value. 
In addition, the signal from the electro-optic detectors leads us to wrong
conclusions about the shape of the original THz pulses.
Figure~\ref{fig:compare_detectors} shows that both ZnTe and GaP crystals
predict additional field oscillations that are absent on the signal measured by
the more precise ABCD method and the Michelson interferometry~\cite{Bakker1998}.
Thus, the electro-optic detection, although it presents very high sensitivity,
strongly affects our perception of the THz electric field deforming its original
shape and misleading about its duration.

In the following sections we demonstrate how one can actually use lossy
channels, like the electro-optic detection, yet still be able to fully recover
the original broadband wave packet. 

\section{The reconstruction method}
In this section we introduce our reconstruction method whose aim is to restore
broadband wave packets that were strongly distorted by a transmission or
detection channel with losses.
The method is based on the nonlinear curve fitting algorithm applied to
experimentally recorded signals.
In this algorithm we choose as initial condition a fitting function that, we
believe, provides the best fit to a series of experimental data points.
The fitting function can depend nonlinearly on a number of parameters, which
are defined by the nonlinear curve fitting algorithm through successive
iterations, starting from some initial estimates.
Usually, the fitting function is expressed in a closed-form, that is, using a
finite number of constants, variables, and certain well-known functions (roots,
exponents, trigonometric functions, etc.).
However, the fitting function can be replaced by any functional relation that
provides a one-to-one correspondence between a set of input data points and data
points to be fitted.
In our reconstruction method we choose a model that represents an input wave
packet and depends on a number of fitting parameters that we want to find (for
example, duration and/or amplitude of the wave packet).
The appropriate model can be found through some prior knowledge about the
process of the wave packet generation or through additional measurements that
are able to measure the wave packet with minimal distortions.
Additionally, we choose a model that encodes the physical properties of the
lossy channel.
This model can also include other fitting parameters like, for example, the
distance of the wave packet propagation and/or the dispersion coefficient.
We use all of the above fitting parameters as the arguments of the fitting
function which takes the model wave packet and simulates its propagation through
the lossy channel.
The output signal, obtained as the result of the simulation, is then compared to
the experimentally recorded signal.
Using the iterative nonlinear curve fitting algorithm we find the values of the
fitting parameters which provide the best match between the simulated and
experimental signals.
As a result, we are able to retrieve the parameters of the input wave packet,
before it was distorted by the lossy channel, and the unknown properties of the
lossy channel itself.
In the rest of this section we consider a specific case of lossy detection of
ultrashort THz pulses by the electro-optic detection scheme.

In order to choose a model for a THz pulse that enters the electro-optic
detection system, let us turn to the results of the ABCD method and Michelson
interferometry (see Fig.~\ref{fig:compare_detectors}).
The signals obtained by these two techniques show that the original THz pulse is
a single cycle pulse, whose electric field $E(t)$ can be modeled with the
following function:
\begin{align} \label{eq:model}
    E(t) = E_0\frac{t}{t_0} \exp\left(-\frac{t^2}{t_0^2}\right),
\end{align}
where $E_0$ is the peak amplitude.
The same expression for $E(t)$ is also predicted by the photocurrent
model~\cite{Kim2007}. 
The details about the model and simulation of electro-optic detection can be
found in the Appendix.
As the duration of $E(t)$ in equation~\eqref{eq:model} we use the value
$\tau_0=2t_0/\sqrt{2}$, which is defined as the time between the first minimum
and the first maximum of the field, similar to the above definition of $\tau_s$.

In our reconstruction method we use two fitting parameters, namely, the duration
$\tau_0$ of the initial THz pulse and the length $d$ of the electro-optic
crystal.
We consider $d$ as an additional fitting parameter in order to compensate errors
rising from inaccurate determination of the crystal's length and cut angle
together with possible misalignment of the crystal relative to the optical
axis.
As the initial estimates for $\tau_0$ and $d$ we use the duration $\tau_s$ of
the recorded electro-optic signal and the length of an electro-optic crystal
provided by its manufacturer.
At this stage of the reconstruction we do not take into account the absolute
value of the amplitude $E_0$ and consider only normalized electro-optic signals.

In practice, the measured electro-optic signal is represented by a sequence of
samples recorded in a finite time window.
The sampling interval and the size of the time window usually differ from the
spacing and size of a numerical grid used in the simulations.
Therefore, before applying our reconstruction method, we prepare the recorded
electro-optic signal.
First, we use the linear interpolation and zero padding in order to transfer the
experimental signal to a numerical grid.
Then, we smooth the result by convolving it with a scaled Hanning window.
We apply this preparation procedure to all electro-optic signals as the first
stage of our reconstruction method.

\begin{figure*}
    \includegraphics{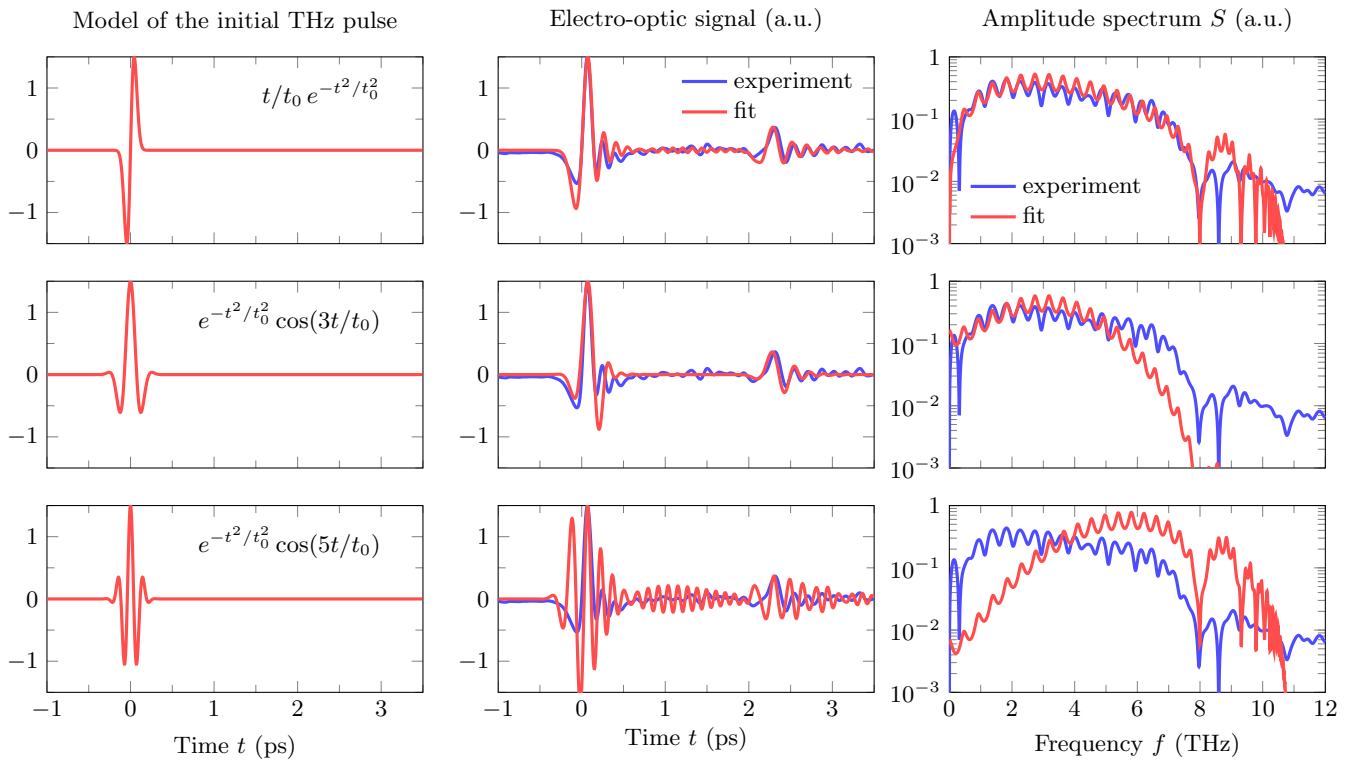}%
    \caption{\label{fig:restore_test}%
             The electro-optic signal measured by 100\,$\mu$m Gap crystal
             (middle column) and its amplitude spectrum (right column) together
             with their best fits obtained by the reconstruction method for
             different models of the initial THz pulse (left column).}
\end{figure*}

In order to test our reconstruction method, let us apply it to the signal
measured by a 100\,$\mu$m GaP crystal.
On Fig.~\ref{fig:restore_test} we plot the experimental electro-optic signal
(middle column) and its amplitude spectrum (right column) together with their
best fits obtained by our reconstruction method for different models of the
initial pulse (left column).
One can clearly see that the model from Eq.~\eqref{eq:model} (top raw) gives
the closest agreement with the experimental findings.
However, the duration $\tau_0$ of the original THz pulse, restored by our
reconstruction method for this model, is equal to 84\,fs.
This value is unreasonably large, since we have already seen that according to
Michelson interferometry the actual duration of the original THz pulse should be
around 44\,fs.
In order to clarify this inconsistency we would like to remind that during
electro-optic detection we lose a lot of information about the high-frequency
part of the THz pulse spectrum.
Therefore, if we consider two very short THz pulses with different durations
such that their spectra are much wider than the detection bandwidth, then their
electro-optic signals will be indistinguishable, because short THz pulses with
the electric field from equation~\eqref{eq:model} have very similar low
frequency spectrum.
In other words, there is no unique solution for our reconstruction method in
case of very short THz pulses.
In practice, the initial estimate for $\tau_0$ is the duration $\tau_s$ of a
recorded electro-optic signal, which is longer than the original THz pulse.
The reconstruction method starts to search successively for shorter durations
and stops at the first one which provides a good fit.
Thus our reconstruction method finds the longest possible THz pulse whose
electro-optic signal fits the recorded one.

Let us make an additional test in order to clarify the amount of error that
our reconstruction method produces during the retrieval of the original THz
pulse duration.
For this purpose let us consider several THz pulses of different durations
$\tau_0$ whose electric fields are calculated by equation~\eqref{eq:model}.
We simulate their electro-optic detection by the 100\,$\mu$m GaP crystal and
then apply our reconstruction method to the obtained results in order to
retrieve the original value of $\tau_0$.
In order to reproduce the experimental conditions, in the simulated
electro-optic signals we filtered out (using a supergauss filter) all
frequencies that are higher than 8\,THz.
Despite the fact that the GaP phonon resonance is located at 11\,THz, in the
experiments we can not see any signal above 8\,THz.
One of the possible explanations is a low signal to noise ratio at these
frequencies.
Additionally, we use a 5\,ps long segment from the numerical grid in order to
reproduce a finite time window which we use in the experiment for sampling the
experimental data.

\begin{figure}
    \includegraphics{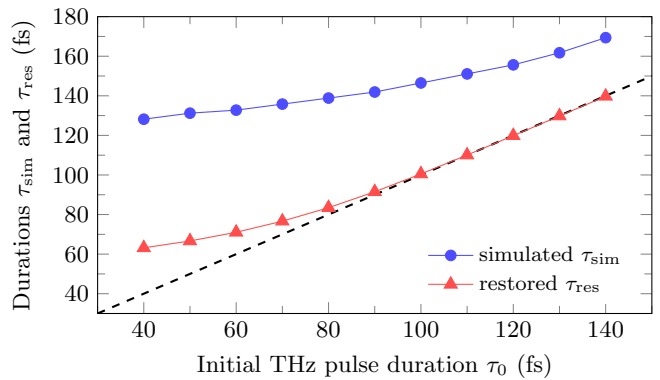}%
    \caption{\label{fig:restore_duration_test} The duration $\tau_\text{sim}$ of
             the simulated electro-optic signal and the corresponding duration
             $\tau_\text{res}$ obtained by the reconstruction method as
             functions of the initial THz pulse duration $\tau_0$.
             Dashed line: see the text.}
\end{figure}

Figure~\ref{fig:restore_duration_test} shows the duration $\tau_\text{sim}$ of
the simulated electro-optic signals and the corresponding duration
$\tau_\text{res}$ restored by our reconstruction method as functions of the
initial duration $\tau_{0}$.
The dashed line in Fig.~\ref{fig:restore_duration_test} indicates an ideal case
when the duration $\tau_\text{res}$ retrieved by the reconstruction method is
equal to $\tau_0$.
Figure~\ref{fig:restore_duration_test} shows that the duration $\tau_\text{sim}$
of the simulated electro-optic signal is always larger than the duration
$\tau_{0}$ of the original THz pulse.
Although for longer THz pulses the difference between $\tau_\text{sim}$ and
$\tau_{0}$ decreases, due to the effects of THz pulse propagation inside the
crystal, such as dispersion, the electro-optic detection still gives too high
values.
In Fig.~\ref{fig:restore_duration_test} we see that for THz pulses whose
duration $\tau_0$ is below than approximately 90\,fs the restored duration
$\tau_\text{res}$ is too high.
In turn, for the THz pulses longer than $90$\,fs, the durations
$\tau_\text{res}$ retrieved by our reconstruction method are exactly equal to 
the initial durations $\tau_0$.
The value of 90\,fs is approximately the inverse of 11\,THz which is the
frequency of the GaP phonon resonance.

The results presented in Fig.~\ref{fig:restore_duration_test} give us a hint
about an additional procedure that will help us to retrieve the original
duration of short THz pulses.
Experimentally we can stretch the generated THz pulses up to the point where the
width of their spectra will be narrower than the detection bandwidth of the
electro-optic crystal (according to Fig.~\ref{fig:restore_duration_test} for
100\,$\mu$m GaP crystal the appropriate durations must exceed 90\,fs).
For such long THz pulses we can apply our reconstruction method in order to
exactly retrieve their original durations.
Then we could use these findings and extrapolate them into the region of shorter
THz pulses, given that there is a known correlation between the widths of the
laser and the generated THz pulses.
In the next section we show that such a correlation indeed exists.

\section{The correlation between the laser and THz pulse durations}
In this section we find the law according to which the duration of the generated
THz pulse depends on the duration of the input laser pulse.
In order to find this law we use the well accepted photocurrent
model~\cite{Kim2007}.
Since in the experiments the stretching of the laser pulse is usually achieved
by changing the distance between the gratings of the laser compressor, the
resulting pulse becomes chirped.
Therefore, as the initial condition for the photocurrent model we use a chirped
two-color pulse with the electric field $E(t)$ given by
\begin{align*}
    E(t) & = E_{\omega_0} A(t) e^{-i\omega_0 t} + E_{2\omega_0} A(t)
             e^{-i2\omega_0 t - i\pi/2}
\end{align*}
and the envelope
\begin{align*}
    A(t) & = \frac{\sqrt{1+iC}}{\sqrt{1+C^2}}
             \exp\left[-(1+iC)\frac{t^2}{2t_f^2}\right],
\end{align*}
where $C$ is the chirp parameter, with $t_f=t_{f0}\sqrt{1+C^2}$ and $t_{f0}$
being the durations of the stretched and transform-limited input laser pulse.
As the central frequency $\omega_0$ we choose the frequency that corresponds to
800\,nm wavelength.
Also we choose the amplitudes $E_{\omega_0}$ and $E_{2\omega_0}$ that correspond
to the intensities $I_{\omega_0}$ and $I_{2\omega_0}$ such that
$I_{2\omega_0}=0.2I_{\omega_0}$.

With the specified initial condition we calculate the photocurrent derivative
$\partial J/\partial t$.
As the generated THz pulse we consider a waveform that corresponds to the
spectrum of $\partial J/\partial t$ at frequencies below 100\,THz.

Since the initial two-color pulse is chirped, we need to take into account that
its phase is strongly modulated.
Therefore, in order to avoid errors in calculations of the amount of free
electrons, we use a phase sensitive nonadiabatic ionization rate proposed by
Yudin and Ivanov~\cite{Yudin2001}.

\begin{figure}
    \includegraphics{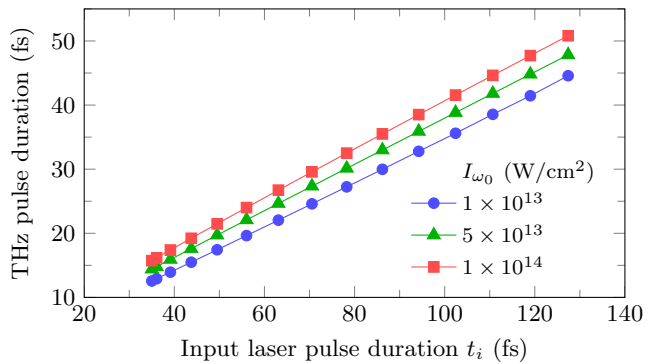}%
    \caption{\label{fig:thz_vs_ir_pulse_duration}%
             The dependence of the THz pulse duration on the duration $t_i$ of
             the input laser pulse, calculated using the photocurrent model.}
\end{figure}

Figure~\ref{fig:thz_vs_ir_pulse_duration} shows the dependence of the duration
of the generated THz pulse as a function of the duration of the input laser
pulse at different intensities $I_{\omega_0}$.
We see that independently of the initial intensity the duration of THz pulse
depends linearly on the duration of the input laser pulse, and the slope of this
linear dependence remains constant.
This finding is of fundamental importance for the successful completion of our
reconstruction process.
Thus for our extrapolation procedure it is now justified to accept that the
duration of the generated THz pulse depends linearly on the duration of the
input laser pulse.

\section{Retrieved duration of the THz pulse}
Now let us apply the combination of our reconstruction method and the
extrapolation procedure to the experimental electro-optic signals.
We conduct a series of experiments with the input laser pulses of different
duration.
For each experiment in the series we measure the FWHM duration $\tau_i$ of the
input laser pulse and detect the corresponding generated THz pulses with a
100\,$\mu$ GaP crystal.
Then for each of the recorded electro-optic signals we apply our reconstruction
method trying to retrieve the duration $\tau_0$ of the original THz pulse.

\begin{figure}
    \includegraphics{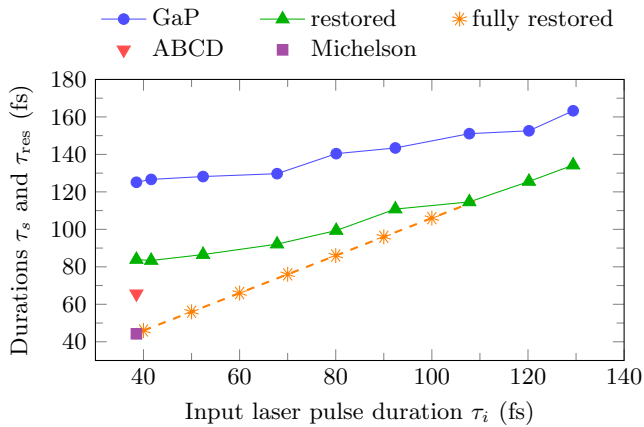}%
    \caption{\label{fig:restore_duration}%
             The duration of the electro-optic signal $\tau_s$ and the
             corresponding duration $\tau_\text{res}$, restored by the
             reconstruction method, as a function of the input laser pulse
             duration $\tau_i$.}
\end{figure}

Figure~\ref{fig:restore_duration} shows the duration $\tau_s$ of the
electro-optic signal measured by the GaP crystal and the corresponding duration
$\tau_0$ retrieved by our reconstruction method as a function of the input laser
pulse duration $\tau_i$.
As a reference data, in this figure we also plot the durations $\tau_s$ measured
by the ABCD method and the Michelson interferometer for the initial (shortest)
laser pulse.
If we compare the durations $\tau_s$ and $\tau_0$ in
Fig.~\ref{fig:restore_duration} to the similar values in
Fig.~\ref{fig:restore_duration_test}, we will find a familiar behavior.
In Fig.~\ref{fig:restore_duration} we see that for the input laser pulses whose
duration $\tau_i$ exceeds 100\,fs, the values of $\tau_0$ retrieved by the
reconstruction method depend linearly on $\tau_i$.
Since in the previous section we have shown that $\tau_i$ depends linearly on
the duration of the generated THz pulses, we can use this data in order to
linearly extrapolate these $\tau_0$ values to the region of $\tau_i$ below
100\,fs (stars in Fig.~\ref{fig:restore_duration}).
As a result of the extrapolation, we find that the initial laser pulse with a
duration $\tau_i=39$\,fs generates THz pulses with the duration $\tau_0=45$\,fs.
This value is almost identical to the duration of 44\,fs measured by Michelson
interferometry.
Therefore, we can conclude that our reconstruction approach allows us to
retrieve the actual duration of the generated THz pulses using only an
electro-optic crystal as the THz detector.

\section{Absolute amplitude of the THz pulse}
In the previous section we have shown how to retrieve the duration of the
original THz pulse distorted by an electro-optic detector.
One of the key assumptions of our reconstruction method is that the electric
field of the original THz pulse can be well described by
equation~\eqref{eq:model}.
The validity of this assumption is proved, on the one hand, by the measurements
performed by the ABCD method and Michelson interferometry (see
Fig.~\ref{fig:compare_detectors}), and on the other hand, by the very good
agreement between the experimental data and the numerical simulations (see
Fig.~\ref{fig:restore_test}).

The shape of the field in equation~\eqref{eq:model} depends only on one
parameter $t_0=\sqrt{2}\tau_0/2$.
Therefore, since we are able to retrieve the duration $\tau_0$, we can retrieve
the exact shape of the original THz pulse.
If, additionally, we find the way to retrieve the absolute amplitude $E_0$, we
will retrieve all the information about the original THz wave packet.

There are two standard methods to calculate the absolute amplitude $E_0$ of a
THz electric field if one uses electro-optic detection.
In order to apply the first method we need to measure the total energy of a THz
pulse.
Additionally, for the same pulse, we need to measure its spatial distribution
and electro-optic signal, then normalize them and integrate.
The desired amplitude $E_0$ can be found if we divide the total THz energy by
the two calculated integrals~\cite{Blanchard2009}.

The second method is based on a simplified version of the theory of
electro-optic detection (see Appendix) where we neglect the dispersion inside
the electro-optic crystal.
As a result of this simplification, we find that the peak of an electro-optic
signal corresponds to a relative phase shift $\Gamma$ given
by~\cite{Casalbuoni2008}
\begin{align} \label{eq:Gamma2}
    \Gamma = \frac{2\pi n_0^3d}{\lambda_0} r_{41}E_0,
\end{align}
where $r_{41}$ is calculated at some specific frequency.
The resulting value of $E_0$ is then calculated by equation~\eqref{eq:signal},
using the readings from the balanced photodiodes of the electro-optic detection
setup.

Both these methods have substantial disadvantages.
In the first method we assume that the temporal distribution of the THz field is
well represented by an electro-optic signal which, as we have already seen, is
not true.
Moreover, this method demands additional measurements of the spatial profile of
the THz pulse.
In turn, the second method does not take into account the dispersion inside an
electro-optic crystal.

As an alternative to the existing methods, we can calculate the absolute
amplitude of the THz electric field using our reconstruction method.
The calculation stages are the following.
First, using a normalized electro-optic signal we restore the duration $\tau_0$
of the initial THz pulse, thereby obtaining its shape, that is, its normalized
electric field.
Then, we simulate the electro-optic signal for a THz pulse with the restored
duration $\tau_0$ and a unit amplitude, let say 1\,kV/cm.
As a result, for a given THz pulse with a given $\tau_0$, we find the relation
between the absolute amplitude $E_0$ (in physical units) and the peak
value of the corresponding electro-optic signal.
The final scaling can be done using equation~\eqref{eq:signal} and the readings
from the balanced photodiodes.
Thus our result will be the closest one to the actual value of the absolute
amplitude, since it does not depend on additional measurements of the pulse
profile and takes into account the distortions introduced by the electro-optic
detector including the dispersion of the electro-optic crystal.

Now, let us determine the absolute amplitude of the THz pulse whose duration we
have retrieved in the previous section.
Following the above procedure, we find that our THz pulse with duration
$\tau_0=45$\,fs has peak amplitude $E_0$ equal to 250\,kV/cm.
For comparison, the amplitude calculated using equation~\eqref{eq:Gamma2} is
equal to 70\,kV/cm, which is 3 times lower than the value obtained by our
reconstruction method.

\section{Conclusions}
In conclusion, we have shown that one can fully retrieve ultrashort broadband
THz wave packets that were severely transformed from dispersive and bandwidth
lossy detection channels, like the very popular electro-optic detection.
For achieving this we introduced a nonlinear curve fitting process in which we
encoded the electro-optic detection physics, while we have also found a way to
recover the lost bandwidth information, performing simple experiments, fully
supported by the photocurrent model of THz generation.
Our results beyond the evident impact on THz science and technology can open the
way in suggesting similar solutions to other problems in physics, optics and
engineering, where only partial information of wave packets and wave functions
can be measured.

\appendix*
\section{Model and simulation of electro-optic detection}
Here we give a brief theoretical description of electro-optic detection and
provide an algorithm for its numerical simulation.
The electro-optic sampling technique is based on measuring the birefringence
induced by THz pulses inside a detection crystal.
To detect this induced birefringence we use a probe laser pulse that changes its
polarization during the propagation through the detection crystal.
After the crystal we split the probe laser pulse into two components with
mutually perpendicular polarizations and send them to a pair of photodiodes.
Let us denote the voltage in each of the photodiodes as $A_1$ and $A_2$.
A normalized difference of $A_1$ and $A_2$ is given by~\cite{Brunken2003}
\begin{align} \label{eq:signal}
    \frac{A_1 - A_2}{A_1 + A_2} = \sin\Gamma,
\end{align}
where $\Gamma$ is a relative phase shift between two orthogonally polarized
components of the probe laser pulse.
We can write $\Gamma$ as~\cite{Casalbuoni2008}
\begin{align} \label{eq:Gamma}
    \Gamma(t) = \frac{2\pi n_0^3d}{\lambda_0} \text{Re}[F(t)],
\end{align}
where $\lambda_0$ is the wavelength of the probe pulse, $n_0$ is the refractive
index of the detection crystal at $\lambda_0$, and $d$ is the crystal thickness.
For GaP crystal at $\lambda_0=800$\,nm we have $n_0=3.18$~\cite{Casalbuoni2008}.
Factor $\text{Re}[F(t)]$ in equation~\eqref{eq:Gamma} denotes the real part of
function $F(t)$ which we define as
\begin{align} \label{eq:F}
    F(t) & = \frac{1}{2\pi} \int\limits_{-\infty}^{\infty}
             H(\omega) \, \widetilde{E}(\omega) e^{-i\omega t} d\omega.
\end{align}
Here $\widetilde{E}(\omega)$ is the spectrum of the incoming THz pulse with the
electric field $E(t)$, that is,
\begin{align} \label{eq:Ew}
    \widetilde{E}(\omega) = \int\limits_{-\infty}^{\infty} E(t)e^{i\omega t} dt
\end{align}
and $H(\omega)$ is an electro-optic response function, given
by~\cite{Casalbuoni2008}
\begin{align} \label{eq:H}
    H(\omega) = r_{41}(\omega) \, G(\omega) \, T(\omega),
\end{align}
where $r_{41}(\omega)$, $G(\omega)$, and $T(\omega)$ are the frequency dependent
electro-optic coefficient, geometric response function, and transmission
coefficient of the detection crystal, respectively.

The transmission coefficient $T(\omega)$ reads as
\begin{align} \label{eq:T}
    T = \frac{t_1t_2e^{ikd}}{1 + r_1r_2e^{2ikd}}.
\end{align}
If we assume that the detection crystal is surrounded by dry air with a unit
refractive index, then $t_1 = 2/(1+n)$, $t_2 = 2n/(n+1)$, and $r_1=(1-n)/(1+n)$,
$r_2=(n-1)/(n+1)$, where $n$ is the refractive index of the detection crystal at
THz frequencies with $k=n\omega/c_0$ being the wave number of the THz pulse.

For the frequency dependent refractive index $n$ we use the following
model~\cite{Casalbuoni2008}:
\begin{align}
    n^2(\omega) = \varepsilon_{el} + \frac{S_0\omega_0^2}
                  {\omega_0^2 - \omega^2 - i\Lambda_0\omega}.
\end{align}
For a GaP crystal we take $\varepsilon_{el}=8.7$, $S_0=1.8$,
$\omega_0/(2\pi)=10.98$\,THz, and
$\Lambda_0/(2\pi)=0.02$\,THz~\cite{Casalbuoni2008}.

The geometric response function $G(\omega)$ takes into account the mismatch
between the phase velocity of the THz pulse and group velocity $v_g$ of the
probe laser pulse inside the detection crystal~\cite{Casalbuoni2008}:
\begin{align}
    G(\omega) & = \frac{1}{d} \int\limits_0^d dz \int\limits_{-\infty}^{\infty} 
                  \delta(z/v_g-t) e^{i(kz-\omega t)} dt \notag \\
              & = \frac{1}{d} \frac{e^{i(k-\omega/v_g)d}-1}{i(k-\omega/v_g)}.
\end{align}
For a probe pulse at the wavelength $\lambda_0=800$\,nm propagating inside a
GaP crystal we have $v_g=0.28c_0$~\cite{Casalbuoni2008}.

The electro-optic coefficient $r_{41}$ determines the sensitivity of a detection
crystal at different frequencies.
We use the following model for $r_{41}$~\cite{Casalbuoni2008}:
\begin{align} \label{eq:r41}
    r_{41}(\omega) = d_E\left(1 + \frac{C\omega_0^2}{\omega_0^2 - \omega^2 -
                                                     i\Lambda_0\omega}\right)
\end{align}
with the same parameters $\omega_0$ and $\Lambda_0$ as in the model for
$n(\omega)$.
For GaP crystals we use $d_E=1.13\times10^{-12}$\,m/V and $C=-0.47$.
These values of $d_E$ and $C$ slightly differ from the ones proposed
in~\cite{Casalbuoni2008}.
We modified the original values in order to shift the frequency where the real
part of $r_{41}$ becomes zero.
This correction gives a better fit for our experimental data.

Equations \eqref{eq:signal}--\eqref{eq:r41} allow us to simulate the process of
electro-optic sampling for any given THz pulse.
Let us assume that we know the electric field $E(t)$ of an initial THz pulse,
then the algorithm of the numerical simulation is the following:
(i) calculate the spectrum $\widetilde{E}(\omega)$ of $E(t)$ in accordance with
equation~\eqref{eq:Ew};
(ii) using equations~\eqref{eq:T}--\eqref{eq:r41}, calculate the electro-optic
coefficient $r_{41}(\omega)$, geometric response function $G(\omega)$, and
transmission coefficient $T(\omega)$;
(iii) calculate the electro-optic response function $H(\omega)$ by
equation~\eqref{eq:H} and then function $F(t)$ by equation~\eqref{eq:F};
(iv) using equation~\eqref{eq:Gamma}, calculate the phase shift $\Gamma(t)$.
The sine of $\Gamma$, according to equation~\eqref{eq:signal}, gives us an
electro-optic signal which we can directly compare with the experimental data.

For a practical realization of the nonlinear curve fitting algorithm we use
\textsc{Python} programming language and specifically the \textsc{curve\_fit}
function from \textsc{scipy.optimize} library~\cite{Python}.
This function implements the Levenburg-Marquardt algorithm for nonlinear curve
fitting.

\begin{acknowledgments}
We gratefully acknowledge the assistance Mrs. Christiana Alexandridi on the
Michelson interferometry data.
This work was partially supported by the EU “Laserlab Europe” and the Aristeia
project “FTERA” (grant no 2570), co-financed by the European Union and Greek
National Funds.
\end{acknowledgments}

%

\end{document}